%% file: main.tex
\documentclass[runningheads,orivec]{llncs}

\usepackage[T1]{fontenc}
\usepackage[save]{silence}
\usepackage{xspace}
\usepackage{booktabs}
\usepackage{cite}
\usepackage{amsmath} 
\usepackage{algorithmic}
\usepackage{graphicx}
\usepackage{textcomp}
\usepackage{xcolor}
\usepackage{balance}
\usepackage{tcolorbox}
\usepackage{subcaption}
\usepackage{url}
\usepackage{makecell} 
\usepackage[capitalise]{cleveref}
\usepackage{xfp}
\newcounter{NumTableEntries}

\usepackage{siunitx}
\newlength{\numwidth}
\newcommand{\statisticsbox}[1]{\makebox[\numwidth][r]{(#1)}}

\usepackage{color}

\usepackage{tikz}
\usetikzlibrary{shapes.geometric, arrows.meta, positioning, calc}

\usepackage[
    colorinlistoftodos,prependcaption,textsize=tiny
]{todonotes}

\makeatletter
\define@key{Gin}{trim-xlabel}[0 40 0 0]{%
  \setkeys{Gin}{trim={#1},clip}%
}
\makeatother

\newcommand\definetool[2]{\newcommand{#1}{{\textsc{#2}}\xspace}}
\definetool{\Scratch}{Scratch}
\definetool{\Whisker}{Whisker}
\definetool{\Neat}{Neat}
\definetool{\Neatest}{Neatest}
\definetool{\Mosa}{Mosa}
\definetool{\Mio}{Mio}
\definetool{\Newsd}{News/d}


\input{results/man_macros}

\input{results/ProjectModelInfo}
\input{results/sprdbmacros}
\input{results/sprdbProjectModelInfo}

\begin{document}

\title{Scenario-Driven Neuroevolution: Using Models to Guide Test Generation for Games}
\titlerunning{Scenario-Driven Neuroevolution}

\author{Gijs van Cuyck\inst{1}%
\and%
Patric Feldmeier\inst{2}%
\and%
Jan Tretmans\inst{1,3}%
\and%
Gordon Fraser\inst{2}%
}

\institute{
    Radboud University,
    Institute iCIS,
    Nijmegen,
    The Netherlands
    \and%
    University of Passau,
    Germany\thanks{Supported by grant BTHA-JC-2024-41 of the Bavarian-Czech Hochschulagentur.}
    \and
    TNO-ESI,
    Eindhoven,
    The Netherlands%
    \thanks{
    This work is part of the project
    \textit{TiCToC - Testing in Times of Continuous Change},
    project nr 17936, part of the research program
    \textit{MasCot - Mastering Complexity},
    which is supported by the Dutch Research Council NWO.}
}

\maketitle

\begin{abstract}
Automatically generating test inputs for games is challenging, as test generators must master the game to reach advanced program states while also ensuring robustness against the heavy program randomisation inherent to games. The test generator \Neatest therefore optimises test suites consisting of neural networks that reach advanced program states and are robust to program randomisation, as they generate test inputs dynamically based on the current program state. \Neatest is a white-box testing approach that aims to generate a network agent for each yet-uncovered statement or branch of the code using neuroevolution. Due to this iterative test generation approach, the algorithm does not scale well to larger programs that may contain thousands of branches. Furthermore, covering every statement or branch in a game often does not correspond to playing the game as intended.
To alleviate these shortcomings, we propose combining \Neatest with a model-based testing approach that allows game testers to define test scenarios via abstract game models. The test generator then no longer optimises networks to reach all branches or statements of a program, but instead trains networks to replicate the concrete desired testing behaviour expressed by the abstract game model. An evaluation on \gamecount{} \Scratch games across varying genres demonstrates that \Neatest, combined with model-based testing, is able to optimise agents that replicate the desired gameplay behaviour defined in the game models while increasing achieved branch coverage by 7\% compared to the traditional code-guided \Neatest approach.

\end{abstract}

\input{sections/1-Introduction}

\input{sections/2-Background}

\input{sections/3-Approach}
\input{sections/4-Evaluation}

\input{sections/5-RelatedWork}
\input{sections/6-Conclusions}

\bibliographystyle{splncs04}
\bibliography{references}

\end{document}

%% file: results/man_macros.tex
\newcommand{\gamecount}{13}

%% file: sections/1-Introduction.tex
\section{Introduction}

In traditional software testing frameworks, tests specify fixed sequences of inputs along with the desired outputs to verify the behaviour of the system under test. However, for systems with non-deterministic behaviour, such as games, defining fixed test inputs is insufficient since the test input sequences cannot adapt to changes in program behaviour, leading to flaky testing behaviour~\cite{gruber_EmpiricalStudyFlaky_2021, feldmeier_NeuroevolutionBasedGenerationTests_2023}. 

\begin{figure}[t]
    \centering
    \begin{subfigure}{0.49\linewidth}
        \includegraphics[width=\linewidth]{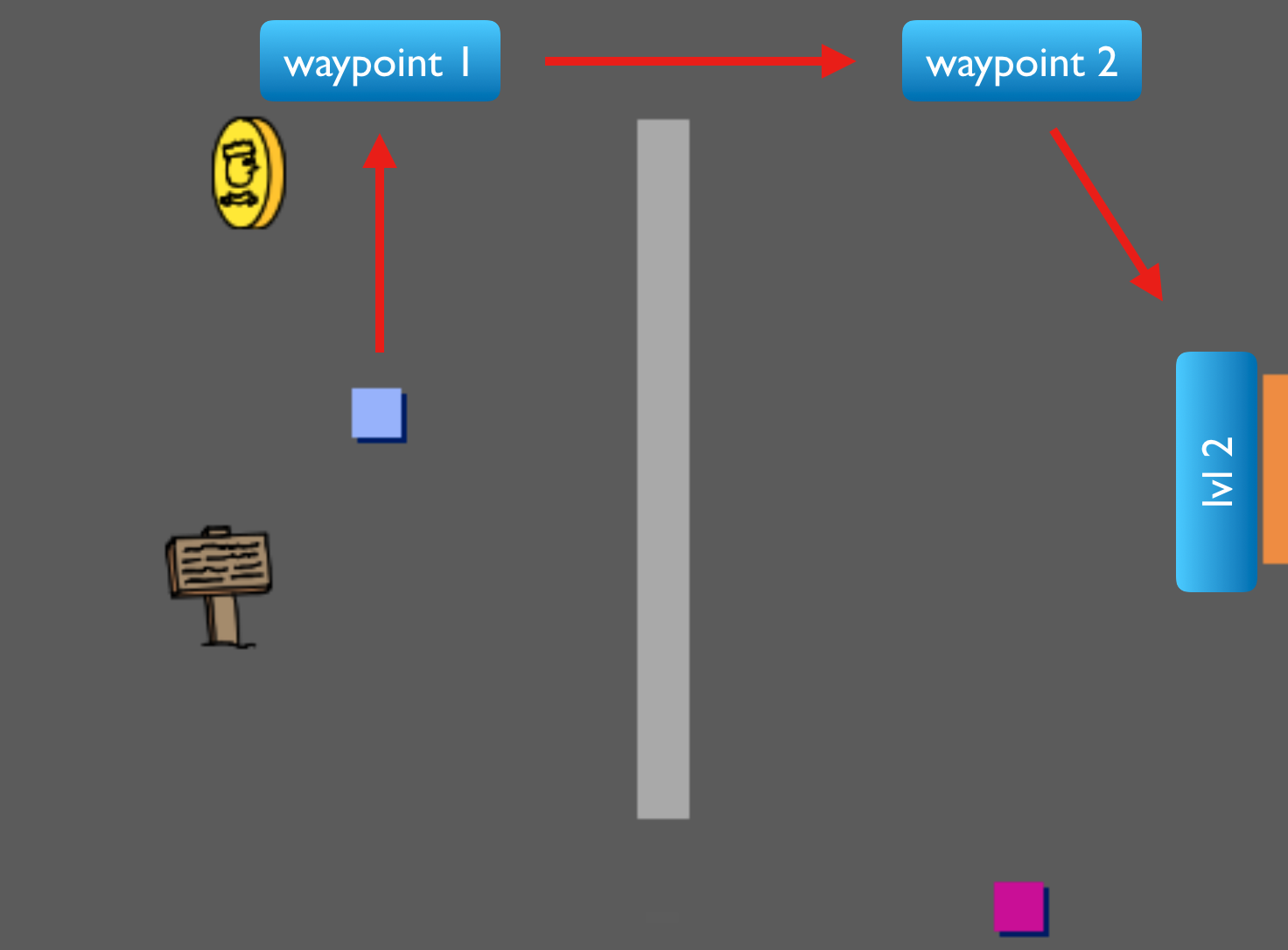}
        \caption{CreateYourWorld}
        \label{fig:createyourownworld}
    \end{subfigure}
    \begin{subfigure}{0.49\linewidth}
        \includegraphics[width=\linewidth]{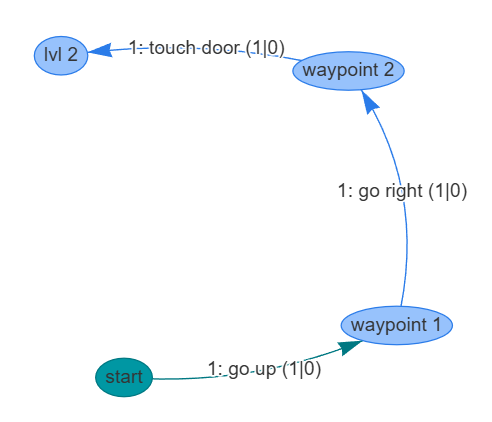}
        \caption{Game model}
        \label{fig:CYW_WP}
    \end{subfigure}
    \caption{The \emph{CreateYourWorld} game and a game model defining the gameplay behaviour to be learnt.}
    \label{fig:scratchgames}
\end{figure}

\Neatest, implemented in the \Whisker~\cite{deiner_AutomatedTestGeneration_2023} testing framework, tackles the challenges of testing non-deterministic systems by employing the evolutionary algorithm \Neat\cite{stanley_EvolvingNeuralNetworks_2002} to generate test suites consisting of neural networks. Since these test suites generate test inputs dynamically based on the current program state, they are robust to program randomisation~\cite{feldmeier_NeuroevolutionBasedGenerationTests_2023}. In order to guide the search, \Neatest treats each branch or statement of a program as a unique optimisation goal and derives a fitness function from each goal that computes the distance towards reaching it.
While the code-guided fitness function generalises to any game, treating each statement or branch of a game as a unique optimisation target of equal importance causes scalability issues, and there may be deceiving fitness landscapes. As an example, consider the \emph{CreateYourWorld} game depicted in~\cref{fig:createyourownworld}, which exposes a deceiving fitness landscape when using \Neatest's default fitness function. In this game, the blue player on the left needs to get to the next level by touching the orange door on the right. \Neatest's code-guided approach will eventually
try to cover a branch or statement that requires it to reach the next level by touching the orange door. For this coverage target, the fitness function computes the distance between the player and the door such that the search can optimise agents to minimise this distance until the player reaches the orange door. However, if direct paths to the exit are blocked by obstacles such as the grey wall, \Neatest tends to get stuck in local optima since the optimised agents must temporarily sacrifice fitness to reach the target area. 

In order to overcome the scalability and local optima issues of \Neatest's code-guided fitness function, we propose to combine \Neatest with a model-based testing approach that allows expressing the desired testing scenario an agent should replicate in the form of a game model. Rather than optimising for every branch or statement in a game, we instruct \Neatest to reach every state of the defined game model and to derive fitness scores based on how close an executed agent came to reaching the currently targeted state.
These game models can be fairly simple, consisting only of a starting state and a single target state, such as reaching the second level in a game. However, we can avoid and overcome problems such as deceptive fitness landscapes by creating more fine-grained models, like the one shown in \cref{fig:CYW_WP}, that define waypoints an agent must reach in sequence to navigate a maze.

Importantly, \Neatest's code-guided algorithm focuses on creating tests that reach specific code locations but fails to differentiate between critical and less important code. A model-guided approach allows a software tester to infer test input generators for specific testing scenarios and program requirements. Building these models requires human effort and transforms \Neatest from a fully automated method to a semi-automated approach. However, the effort required to build these models is limited, as most models are simple to build and can even be reused across games due to them being small and high-level. Linking tests to testing scenarios or program requirements provides better control and insight into what is being tested and yields more actionable feedback when triaging a failing test. Finally, since our approach proposes defining the testing scenario via abstracted game models, a tester does not have to interfere with the source code of the application under test. We believe this approach generalises beyond just testing games and could also be used to add coverage-driven testing to existing model-based testing approaches, which is a hard problem that typically requires the use of constraint solvers, such as SMT \cite{vandenbos_CoverageBasedTestingSymbolic_2019}.

The performance of \Neatest is also highly dependent on the used hyperparameters, for which optimal values are often setting-dependent and hard to find \cite{bischl_HyperparameterOptimizationFoundations_2023}. Using a model as a hyperparameter gives a much more understandable handle for a tester to tweak the coverage direction and performance.

In detail, the contributions of this paper are as follows:
\begin{itemize}
    \item A black-box test generation algorithm using a model-guided fitness function, enabling developers to define test scenarios or requirements to be tested.
    \item An implementation of the model-based fitness function as an extension of the \Neatest algorithm in the \Whisker testing framework.
    \item Two game-agnostic modelling strategies that improve the effectiveness of the evolutionary algorithm.
    \item An empirical evaluation of the proposed approach on a set of \gamecount{} \Scratch games with varying game genres.
\end{itemize}
Although our evaluation uses a diverse set of \gamecount{} \Scratch games~\cite{maloney_ScratchProgrammingLanguage_2010}, the proposed approach is  not limited to \Scratch games but generalises to other game frameworks and non-deterministic applications in principle. In order to be applicable, we only need a model that describes desirable states of the system under test and a test harness that exposes relevant parts of the system's runtime state. For a game engine such as \Scratch or Unity, this test harness can be game-agnostic. Our experiments demonstrate that the model-guided approach successfully optimises agents that can infer actions to execute specific testing scenarios and can even improve branch coverage compared to a white-box approach by 7\%, without directly optimising for coverage.

%% file: sections/2-Background.tex
\section{Background}
We combine neuroevolution with model-based testing to guide the \Neatest algorithm towards generating tests for games that are implemented in \Scratch.

\subsection{The Scratch Programming Environment}
\Scratch~\cite{maloney_ScratchProgrammingLanguage_2010} is an introductory block-based programming environment that is among the most popular educational programming languages, with 135 million registered users.\footnote{June 2026: \url{https://scratch.mit.edu/statistics/}} Creating \Scratch games is similar to developing games in mature game engines: Users start by placing actors, also called sprites, on a 2D canvas whose behaviour is defined by assigning programming \emph{scripts} to them. Scripts are developed by arranging blocks of code in a meaningful order. Each block corresponds to several lines of code in mature programming languages and encodes the program behaviour. Many of these blocks provide commonly required game functionalities, such as user input handlers, changing the visual appearance of sprites, and checking whether game objects overlap. The execution of \Scratch programs resembles the execution flow of other game frameworks by having a central game loop that periodically updates the state of the program based on all currently active scripts. Scripts are activated via specific events, such as user inputs like button presses, or signals that are sent programmatically.

Although the block-based nature of \Scratch avoids syntax errors, flaws in program logic are still possible and frequent. Previous work has therefore introduced the \Whisker testing framework~\cite{deiner_AutomatedTestGeneration_2023}, which allows users to send test inputs to a \Scratch program under test and to validate the observed program behaviour. Tests can be written manually in \emph{JavaScript}, by arranging specialised testing blocks within the \Scratch user interface~\cite{feldmeier_BlockBasedTestingFramework_2024} or by creating model-based tests~\cite{gotz_ModelbasedTestingScratch_2022} in \Whisker's user interface. Moreover, tests can also be generated automatically using \emph{Search-Based Software Testing}~\cite{mcminn_SearchBasedSoftwareTesting_2011} techniques, which are commonly used to create unit tests for applications other than games. However, recent work has shown that these techniques are not suited to reach advanced program states in games and fail to adapt to the inherent randomisation present in games~\cite{deiner_AutomatedTestGeneration_2023, feldmeier_NeuroevolutionBasedGenerationTests_2023}. Thus, recent work has extended \Whisker with a neural network-based testing approach that is robust to randomised program behaviour~\cite{feldmeier_NeuroevolutionBasedGenerationTests_2023}.

\subsection{Neuroevolution-based Testing of Games}
\label{sec:Neatest}

The test generator \Neatest~\cite{feldmeier_NeuroevolutionBasedGenerationTests_2023}, implemented on top of \Whisker, combines \emph{Search-Based Software Testing} with neuroevolution to generate 
neural network agents, where each agent is optimised to produce user inputs that reach specific parts of the game regardless of randomised program behaviour.

\Neatest is an iterative test generation algorithm, with each iteration starting by selecting one coverage objective, which corresponds to reaching a program statement or branch in the game under test. With the objective set, \Neatest employs the neuroevolution algorithm \Neat~\cite{stanley_EvolvingNeuralNetworks_2002} to optimise network agents towards producing game inputs that achieve the selected objective reliably. To this end, the evolutionary algorithm generates an initial population of networks consisting solely of the input and output layers whose nodes are fully connected with randomised weights.
Next, the algorithm assigns each network a fitness score that reflects how close the agent was to solving the objective by letting the agent interact with the game under test.
\Neatest uses a coverage-guided fitness function that is defined as a linear combination of the approach level (AL) and the normalised branch distance (BD): $f = \text{AL} + \alpha(\text{BD})$, with $\alpha$ representing the normalisation function $\alpha(x) = x / (x + 1)$~\cite{arcuri_ItDoesMatter_2010}. The approach level corresponds to the number of control dependencies between the immediate control dependency of the coverage objective and the closest covered objective in the execution trace, whereas branch distance indicates the distance towards satisfying the first missed control dependency. \Neat continues by selecting the best networks based on their fitness scores and generating a new generation of networks by modifying the selected networks via crossover and mutation. Crossover combines the weights of two parents, while mutation randomly perturbs connection weights and the networks' topology. Over many such generations, \Neat co-evolves the topology and weights of networks to solve the given task. Once a network that reaches a targeted statement or branch is found, \Neatest re-executes the given network multiple times using varying random number generator seeds to verify that the network is capable of reaching the targeted objective reliably, even in the face of strong program randomisation. Once a robust network is found, it is stored for later retrieval, and \Neatest continues with the next coverage objective until a predefined search budget is exhausted or all objectives are reached.

\subsection{Model-Based Testing}
Model-based testing uses models of the system under test to generate tests. 
Our game models are a variant of extended finite state machines~$(S, q_0,P, \Sigma, E)$:
\begin{itemize}
    \item $S$ denotes a finite set of abstract model states.
    \item $q_0 \in S$ denotes the initial state of the game model.
    \item $P$ is a set of propositions over the game state.
    \item $\Sigma$ is a set of guards, which are Boolean functions over $P$.
    \item $E \subseteq S \times \Sigma  \times S$ denotes a finite set of transitions labelled with guards.
\end{itemize}

The states of the game model represent abstractions of the possible states the tested system can be in. Edges represent ways of going from one state to another, which can only be taken if their associated guard evaluates to true. Guards are boolean expressions that are evaluated on the program state $P$. $P$ represents an abstraction over the real state of the program, where only relevant aspects are made visible to the model and testing framework. This is also known as a test harness. In our application scenario of testing \Scratch games, the program state corresponds to the state of the \Scratch virtual machine at runtime. Thus, $P$ includes positions, rotation, costumes, variables and the presence of program events, such as text output by sprites or whether two sprites are touching.

Inspired by the concept of test purposes from the field of model-based testing~\cite{devries_FormalTestPurposes_2001}, we create one or more small models of the tested game's behaviour, instead of a single general-purpose model that covers everything. In this paper, we focus on input generation, but the \Whisker framework also supports adding effects to edges, which are boolean conditions that are evaluated when an edge is taken. These effects can then be used as test oracles.

%% file: sections/3-Approach.tex
\section{Inferring Actions from Game Models through Neuroevolution}

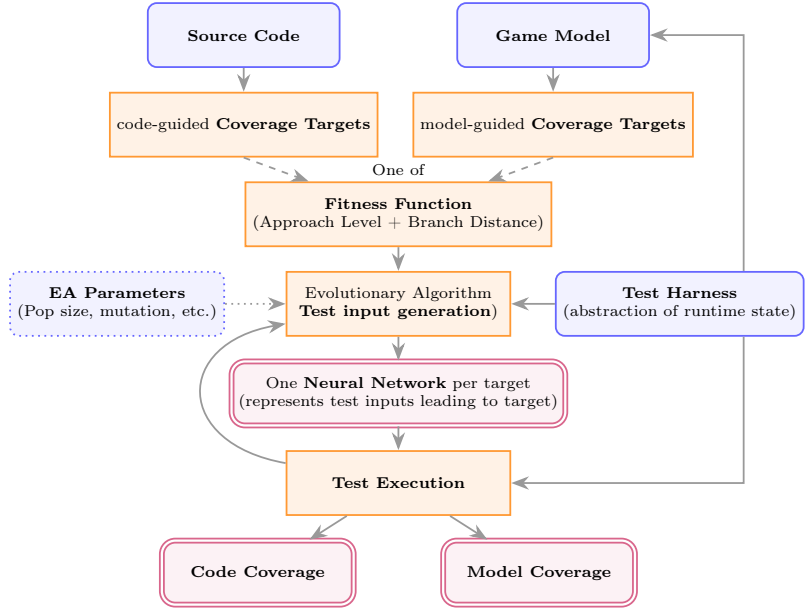
\begin{figure}[tb]
    \centering
    \resizebox{.9\linewidth}{!}{\input{images/process-diagram}}
    \caption{An overview of the testing process. Rounded nodes are inputs, and double-lined nodes are outputs. Dotted nodes and lines are out of scope.}
    \label{fig:testingOverview}
    \vspace{-1em}
\end{figure}

\Neatest is an evolutionary algorithm which iteratively generates a population of candidate solutions in the form of neural networks that are optimised for reaching a specific coverage target.
As shown in~\cref{fig:testingOverview}, we propose deriving these coverage targets from a game model instead of directly from the source code.
The resulting optimised networks are trained to generate inputs that reliably reach their targeted model state from the initial state, which represents the start of the game. To evaluate how close a network gets to solving its task, we execute the network on the game under test and record the resulting execution trace. We then compute the fitness function similar to the code-guided fitness function described in~\cref{sec:Neatest}.
However, our model-guided approach represents control locations as model states $S$ and derives the branch distance from the model's edge guards $\Sigma$.
Thus, the approach level is calculated as the number of states between the target state and the closest state reached during test execution, whereas branch distance measures the distance towards turning the guard on the relevant edge that leads closer to the target state to true.

For instance, considering the model in \cref{fig:CYW_WP}, with the current goal set to reach the second level, and given that an agent has already reached the first waypoint but not the second, the approach level would be 1. The branch distance is then computed based on the guard of the edge linking the two waypoints.

The branch distance for guards, such as the ones in~\cref{fig:CYW_WP}, that evaluate whether some location in the game was reached or if two sprites touch each other, is calculated by computing the distance between the player and the target.
Guards that compare variables or time compute the absolute difference between the real and expected values as the branch distance. For other guards, such as whether a sound is playing or a button is pressed, there is no way of knowing how close the guard is to evaluating to true. In these cases, the values 0 for true and 1 for false are used, which do not give the search any extra guidance on how close it is to reaching a given program state. If a guard consists of multiple conditions that must all be true at the same time, their branch distances are summed. If multiple edges are leading to the targeted state, the one on the shortest path is used. In case of a tie, the edge with the lowest branch distance is used.

In this paper, we focus on input data generation for use in testing and leave the remainder of the \Neatest framework, shown in~\cref{fig:testingOverview}, untouched. After executing our generated test suite, we obtain the achieved code and model coverage. However, to obtain a pass/fail verdict, a test oracle is needed, for which several options are already supported by the \Whisker testing framework~\cite{deiner_AutomatedTestGeneration_2023, feldmeier_NeuroevolutionBasedGenerationTests_2023, gotz_ModelbasedTestingScratch_2022}.

%% file: images/process-diagram.tex

\newlength{\xdist}
\newlength{\ydist}
\setlength{\xdist}{4.4cm}
\setlength{\ydist}{1.4cm}
\begin{tikzpicture}[
    node distance=\ydist and \xdist,
    on grid,
    every node/.append style={font=\scriptsize},
    input/.style={rectangle, rounded corners, draw=blue!60, fill=blue!5, thick, minimum width=3cm, minimum height=1cm, align=center},
    process/.style={rectangle, draw=orange!80, fill=orange!10, thick, minimum width=3.5cm, minimum height=1cm, align=center},
    output/.style={rectangle, rounded corners=6pt, draw=purple!60, fill=purple!5, thick, minimum width=3cm, minimum height=1cm, align=center,double, double distance=1pt},
    arrow/.style={-{Stealth[scale=1.2]}, thick, draw=gray!80},
    one of arrow/.style={arrow,dashed},
    optional/.style={dotted},
    opt arrow/.style={arrow,optional},
]

    \node[input] (src) {\textbf{Source Code}};
    \node[input] (model) [right=1.1\xdist of src] {\textbf{Game Model}};
    \node[process] (modeltarget) [below= of model]{model-guided \textbf{Coverage Targets}};
    \node[process] (sourcetarget) [below= of src]{code-guided \textbf{Coverage Targets}};
    \node[process] (fitness) [below right=\ydist and .55\xdist of sourcetarget] {\textbf{Fitness Function}\\ (Approach Level + Branch Distance)};
    \node[] (one of) [above=.5\ydist of fitness] {One of};

    \node[process] (ga) [below=of fitness] {Evolutionary Algorithm \\\textbf{Test input generation})};
    \node[input,optional] (ga-params) [left=\xdist of ga] {\textbf{EA Parameters}\\ (Pop size, mutation, etc.)};
    \node[input] (sut) [right= of ga] {\textbf{Test Harness}\\(abstraction of runtime state)};

    \node[output] (nn) [below=of ga] {One \textbf{Neural Network} per target\\ (represents test inputs leading to target)};

    \node[process] (test-exec) [below=of nn] {\textbf{Test Execution}};


    \node[output] (codecov) [below=of test-exec, xshift=-.5\xdist] {\textbf{Code Coverage}};
    \node[output] (modelcov) [below=of test-exec, xshift=.5\xdist] {\textbf{Model Coverage}};

    
    \draw[arrow] (src) -- (sourcetarget);
    \draw[arrow]    ([xshift=1cm]sut.north) |- (model);
    \draw[arrow] (model) -- (modeltarget);
    \draw[one of arrow] (modeltarget.south) -- (fitness);
    \draw[one of arrow] (sourcetarget.south) -- (fitness);
    
    \draw[arrow]    (fitness) -- (ga);
    \draw[opt arrow]    (ga-params) -- (ga);
    \draw[arrow]    (sut) -- 
                            (ga);
    \draw[arrow]    (ga) -- (nn);
    \draw[arrow]    (nn) -- (test-exec);
    
    \draw[arrow]    ([xshift=1cm]sut.south) |- (test-exec);
    \draw[arrow]    (test-exec) -- (codecov);
    \draw[arrow]    (test-exec) -- (modelcov);
    \draw[arrow]    (test-exec.170) to[out=170, in=190, looseness=2.1] (ga.190);

\end{tikzpicture}

%% file: sections/4-Evaluation.tex
\section{Evaluation}
\begin{figure}[tbh]
    \begin{subfigure}{0.49\columnwidth}
        \includegraphics[width=\linewidth]{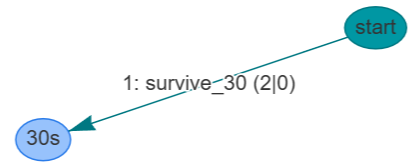}
                \caption{Simple model}
                \label{fig:simple_model}
    \end{subfigure}
    \begin{subfigure}{0.49\columnwidth}
        \includegraphics[width=\linewidth]{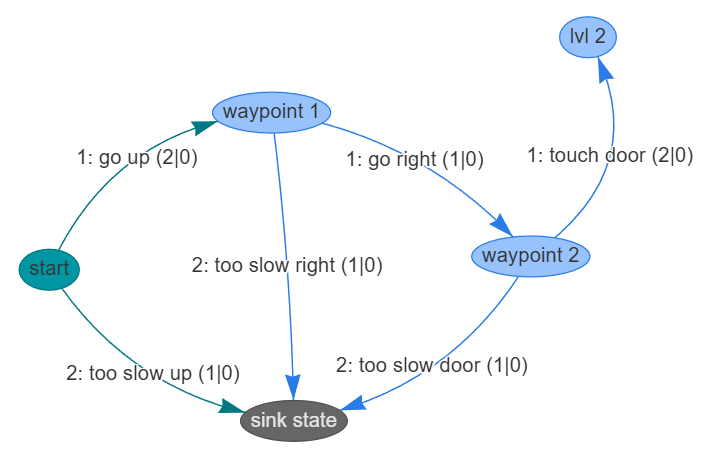}
        \caption{Specialised Model}
        \label{fig:sink_chain_model}
    \end{subfigure}
    \caption{A simple model for surviving 30 seconds in any game and a specialised model with a sink state for \emph{CreateYourWorld}}
    \label{fig:modeltypes}
\end{figure}

To evaluate game models for guiding \Neatest towards useful input generation, we consider two types of models, as shown in \cref{fig:modeltypes}: small, simple models that encode game-agnostic goals, and larger, more specialised models that encode goals for a specific game. We use these to answer the following research questions:
\begin{description}
    \item [RQ1:] How effective are simple models at guiding test input generation?
    \item [RQ2:] Can test input generation be further improved by specialising the models?
\end{description}

\subsection{Dataset}
We establish a dataset of \gamecount{} games for evaluating the proposed approach by combining two existing datasets previously used to assess \Neatest~\cite{feldmeier_LearningViewingGenerating_2023, feldmeier_ManyObjectiveNeuroevolutionTesting_2025}. All dataset games are non-trivial to learn, require meaningful gameplay to win the game, and can be completed in less than a minute of non-accelerated gameplay, given a successful gameplay strategy. Overall, the dataset consists of games with varying genres and levels of complexity as shown in~\cref{tab:project_data}. The \emph{Block} and \emph{Branch} columns show the number of code blocks used to implement the game, and the total number of branches for measuring branch coverage, respectively. Please note that since \Scratch is a domain-specific language for developing games, sophisticated games can already be realised with relatively few blocks.

\begin{table}[t]

    \newcommand{\createTableRow}[4][]{#2 &%
    \csname #1ProjStatements#2\endcsname &%
    \csname #1ProjBranches#2\endcsname %
    }
    \newcommand{\headerline}[0]{
    Project & \multicolumn{1}{c}{Block} & \multicolumn{1}{c}{Branch}
    }
    \centering
    \caption{The evaluation dataset of \gamecount{} \Scratch games.}
    \resizebox{\linewidth}{!}{
    \begin{tabular}{l r r@{\hspace{3pt}}|@{\hspace{3pt}}l r r@{\hspace{3pt}}|@{\hspace{3pt}}l r r }
        \toprule
        \headerline  & \headerline & \headerline \\
        \midrule
        \createTableRow{CatchTheDots}{PlayThirtySecsimple}{Time}&
            \createTableRow{FinalFight}{Win}{Score}&
            \createTableRow{HackAttack}{ScoreThirtySimple}{Score}\\
        \createTableRow{CreateYourWorld}{Simple}{Level}&
            \createTableRow{FlappyParrot}{ScoreOne}{Score}&
            \createTableRow[sprdb]{Pong}{SurviveFifteenSec}{Time}\\
        \createTableRow{Dodgeball}{Simple}{Position}&
            \createTableRow{Frogger}{Eatfly}{Position}&
            \createTableRow{Snake}{Scoreone}{Score}\\
        \createTableRow{Dragons}{Playfifteensecssimple}{Time}&
            \createTableRow{FruitCatcher}{SurviveThirtySecSimple}{Time}&
            \createTableRow[sprdb]{SpaceOdyssey}{PlayThirtySec}{Time}\\
        \createTableRow{WhackAMole}{ScoreThirtysimple}{Score}\\
        \bottomrule
    \end{tabular}
    }
    \label{tab:project_data}
    \vspace{-1em}
\end{table}

\subsection{Experimental Setup}
All experiments are executed on a computing cluster consisting of 7 nodes, each equipped with 2 AMD EPYC 9554 CPUs running at 3.10GHz. We implemented our model-based approach as a fitness function, which can be configured via a hyperparameter in the \Whisker testing framework.
All other hyperparameters of the \Neat algorithm are set based on prior work, which has shown that they yield good results across a variety of games~\cite{feldmeier_NeuroevolutionBasedGenerationTests_2023}.
The default \Neatest algorithm switches the currently optimised objective after 10 generations without improvement to avoid wasting search time on infeasible objectives while simpler ones are still within reach. Since in our model-guided approach the objectives are designed by the tester via the game model, we assume that all objectives are feasible to solve. Thus, to not distort the testers' intention expressed through the game model, we do not switch model-based objectives prematurely.

To assess how well the model-based fitness function guides the search towards reaching model states, we compute the achieved state coverage on the models from which the fitness function is derived. Furthermore, we compare the proposed approach with the default code-guided baseline by evaluating the achieved branch coverage and the number of times a game is won.
As in previous work~\cite{feldmeier_NeuroevolutionBasedGenerationTests_2023}, we define a game to be won if a statement corresponding to actually winning the game or reaching an advanced program state requiring meaningful gameplay is covered. To determine statistical significance, we use the \emph{Mann-Whitney-U} test~\cite{mann_TestWhetherOne_1947} with a significance threshold of 0.05. Furthermore, we compute the effect size of the obtained results between the code-guided and the model-guided approaches using the \emph{Vargha-Delaney} $\hat{A}_{12}$ metric~\cite{vargha_CritiqueImprovementCL_2000}, which quantifies the probability that a single execution of the model-guided approach outperforms a single execution of the code-guided baseline.
Due to the stochastic nature of the evaluated games and the \Neatest algorithm, we repeat each experiment 30 times. Similar to the default \Neatest framework, an objective is only considered covered if an agent is also able to solve the same objective in 10 randomised game executions. We set the search budget for all evaluated methods to 5 hours, since initial experiments showed that, after this time limit, the search converges without further significant improvements in branch coverage.

\textbf{RQ1 (Simple Models):}
For an initial investigation on the capability of the proposed approach, we created small game models, such as the one depicted in~\cref{fig:simple_model}, consisting of two to four states that require the agents to learn to play the game meaningfully.
As shown in~\cref{tab:model_data}, the game models encode simple scenarios like surviving for 30 seconds,
reaching the next level, or scoring points.
Since the starting state is trivially covered by starting the game,
we do not count it towards model coverage. Thus, most models in RQ1 encode only one optimisation objective.
Due to the high-level abstraction of the game models, they are mostly game-agnostic. Although the gameplay of both games differ, a model to guide the search towards optimising agents to play \emph{Pong} for 15 seconds can also be used to train agents to play the \emph{Dragons} game, thereby keeping the human effort required to design models low.

\begin{table}[t]
    \newcommand{\createTableRow}[4][]{#2 %
    & \csname #2#3States\endcsname %
    & \csname #2#3Edges\endcsname %
    & #4\\}
    \newcommand{\createTableHeader}[0]{Project & \multicolumn{3}{l}{States  Edges  Scenario}}
    \centering
    \caption{Number of states, edges and defined scenario of every game model used in RQ1 (top) and RQ2 (bottom).}
    \resizebox{\linewidth}{!}{
    \begin{tabular}[t]{lrrl@{\hspace{3pt}}|@{\hspace{3pt}}}
    \toprule
    \createTableHeader\\
    \midrule
    \createTableRow{CatchTheDots}{PlayThirtySecsimple}{Play for 30 seconds}
    \createTableRow{CreateYourWorld}{Simple}{Finish levels 1-3}
    \createTableRow{Dodgeball}{Simple}{Reach end of floors 1-3}
    \createTableRow{Dragons}{Playfifteensecssimple}{Play for fifteen seconds}
    \createTableRow{FinalFight}{Win}{Reduce opponents health}
    \createTableRow{FlappyParrot}{ScoreOne}{Score a point}
    \createTableRow{Frogger}{Eatfly}{Touch the fly}
    \createTableRow{FruitCatcher}{SurviveThirtySecSimple}{Play 30 seconds}
    \createTableRow{HackAttack}{ScoreThirtySimple}{Play 30 seconds}
    \createTableRow[sprdb]{Pong}{SurviveFifteenSec}{Play 15 seconds}
    \createTableRow{Snake}{Scoreone}{Touch an apple}
    \createTableRow[sprdb]{SpaceOdyssey}{PlayThirtySec}{Play 30 seconds}
    \createTableRow{WhackAMole}{ScoreThirtysimple}{Score 30 points}
    \bottomrule
\end{tabular}%
\begin{tabular}[t]{lrrl}
    \toprule
    \createTableHeader\\
    \midrule
    \createTableRow{CatchTheDots}{PlaySixtySec}{Play for 10, 20, 30, 45, 60 seconds}
    \createTableRow{CreateYourWorld}{Es}{\makecell[tl]{Reach next waypoint within\\ 2 seconds}}
    \createTableRow{CreateYourWorld}{EsCoverage}{\makecell[tl]{Reach the waypoints and also\\ touch a coin, sign and person}}
    \createTableRow{Dodgeball}{Dodgeball}{\makecell[tl]{Reach the waypoints\\ while avoiding the balls}}
    \createTableRow{Dragons}{ScoreSixLongchain}{Score a point 6 times}
    \createTableRow{FruitCatcher}{ChainedSurvive}{Play for 3, 6, 8, 12, 15, 30 seconds}
    \createTableRow{FlappyParrot}{ScoreSixLongchain}{Score a point 6 times}
    \createTableRow{SpaceOdyssey}{CatchSevennolose}{Touch a star seven times}
    \createTableRow{WhackAMole}{ScoreThirty}{Score 1, 4, 6, 8, 10, 20, 30 points}
    \bottomrule
\end{tabular}
}
\vspace{-1em}
    \label{tab:model_data}
\end{table}

\textbf{RQ2 (Specialised Models):}
While RQ1 assesses the search efficiency if minimal effort is put into designing the game models, RQ2 investigates whether more fine-grained modelling approaches that promise better guidance for the search can improve the optimisation process. To this end, we adopt an approach similar to \emph{Curriculum Learning}~\cite{bengio_CurriculumLearning_2009} in which we task the search with optimising agents for increasingly difficult objectives. Curriculum Learning tasks agents with increasingly difficult objectives, which serve as stepping stones towards the overall objective. We make use of this technique by creating models that do not task agents with reaching the targeted objective immediately but instead add intermediate states as stepping stones towards the actual objective. For instance, we extend the \emph{FruitCatcher} model, which consists of a single state that tasks the agent with surviving for 30 seconds, with intermediate states that require the agent to survive for increasing steps of 3 seconds each. The aim of \emph{FruitCatcher} is to catch falling apples for 30 seconds, and the game ends if an apple touches the ground. The value of 3 seconds was chosen because it roughly matches the time needed to catch an apple once. Since \Neatest only treats an objective as covered if an agent manages to do so in several randomised trials, the agents can only progress in the model when they learn to catch apples in sequence reliably.

Another way to apply \emph{Curriculum Learning} is to model waypoints agents have to reach within their playthrough. In the maze game \emph{CreateYourWorld} (\emph{CYW}), we extend the models by creating a sequence of states that serve as waypoints guiding the player through the maze. These waypoints are located at the top of the level before the obstacle, at the top of the level after the obstacle and at the orange door leading to the next level, as depicted in \cref{fig:createyourownworld}. This placement creates a smoother learning curve by presenting obstacles one at a time and changing the fitness function when new challenges are introduced.

In addition to \emph{Curriculum Learning}, we extend game models with \emph{sink states}, as it often becomes apparent that an agent will not perform well long before its playthrough ends. A sink state has no outgoing edges, rendering it impossible to reach any other state from a sink state. Thus, agent executions can be stopped prematurely if they enter a sink state, since an improvement of the fitness score is no longer possible. This corresponds to the concept of a `miss' in a test purpose. By stopping playthroughs early, less time is spent on poorly performing agents, allowing the search to evolve more generations within the same search budget. An example of a model that implements \emph{Curriculum Learning} and the sink-state technique is depicted in~\cref{fig:sink_chain_model}.

One of the biggest advantages of using models is that the tester has more control over which parts of the intended game behaviour get tested compared to an automatic approach, which treats all code as equally important.
To better quantify the coverage of user-defined sections of code, we executed the generated tests and measured the average state coverage achieved by the generated tests for several models, compared to the tests generated without models. To limit the experiment runtime, we selected 10 random tests for both the code-guided and model-guided approaches from the set of 30 generated tests used in the previous analysis for each model-game combination. Each test was then executed 30 times to account for randomness.

\subsection{Threats to Validity}
\noindent\textbf{Internal validity:} The inherent randomness of games and neuroevolution may yield different results when experiments are rerun. To account for this stochasticity, we repeated each experiment 30 times and report statistical significance.

\noindent\textbf{External Validity:} While our dataset is small, it encompasses games that require the test generator to learn to play games across different genres with varying controls. Although we used \Scratch games as an evaluation dataset, the proposed approach is not limited to \Scratch and generalises to other platforms. However, we cannot guarantee that our results generalise beyond this dataset.

\noindent\textbf{Construct Validity:}
We use branch coverage as an evaluation metric because bugs can only be triggered when buggy code is executed. However, not all branches are hard to reach, as some might be covered even by randomised inputs. To account for this, we also show the number of wins reached as a measure of how often the compared methods learned to play the game meaningfully. By measuring state coverage on a per-state basis, we can also evaluate in more detail which parts of the game behaviour are or are not covered.

\subsection{RQ1 (Simple Models)}
\label{sec:RQ1}

The results of running \Neatest with simple models compared to code-guided \Neatest are shown in \cref{tab:basic_results}.
On average across all \gamecount{} games, the code-guided approach achieves a slightly higher branch coverage of 85\% compared to 80\% for the model-guided approach. The difference in coverage mainly results from the simple models focusing on reaching the end of a game while ignoring easy-to-reach branches that are not required to win. For most games, there are more of these side branches than there is code that can only be reached by playing the game well, and so the average coverage goes down. For specific games where the only way to increase coverage is to play the game well, such as \emph{CreateYourWorld} or \emph{FruitCatcher}, coverage increases when using game models.
Of the \gamecount{} games in \cref{tab:basic_results},
the simple models yield significantly higher coverage in five games and significantly lower coverage in seven other games. Thus, even though the model-guided approach does not directly optimise for branch coverage, it can still achieve comparable, and sometimes even better, coverage than optimising for branch coverage directly.

\begin{table}[t]
    \newcommand{\BcovcodeSum}{0}
    \newcommand{\BcovmodelSum}{0}
    \newcommand{\codeWinsSum}{0}
    \newcommand{\modelWinsSum}{0}
    \newcommand{\mcovSum}{0}
    \newcommand{\createTableRow}[3][]{%
    \xdef\BcovcodeSum{\fpeval{\BcovcodeSum + \csname #1MeanBranchCoverage#2NeatestBranchFiveHNA\endcsname}}%
    \xdef\BcovmodelSum{\fpeval{\BcovmodelSum + \csname #1MeanBranchCoverage#2StatecoverageFiveH#3\endcsname}}%
    \xdef\codeWinsSum{\fpeval{\codeWinsSum +\csname #1Wins#2NeatestBranchFiveH\endcsname }}%
    \xdef\modelWinsSum{\fpeval{\modelWinsSum +\csname #1Wins#2#3StatecoverageFiveH\endcsname }}%
    \xdef\mcovSum{\fpeval{\mcovSum +\csname #1MeanObjectiveCoverage#2StatecoverageFiveH#3\endcsname }}%
    \stepcounter{NumTableEntries}%
    #2 &%
    \num{\csname #1MeanBranchCoverage#2NeatestBranchFiveHNA\endcsname} &%
    \num{\csname #1MeanBranchCoverage#2StatecoverageFiveH#3\endcsname}\;\statisticsbox{\csname #1EffectSizeBranchCoverage#2NeatestBranchFiveHStatecoverageFiveH#3\endcsname} &%
    \csname #1Wins#2NeatestBranchFiveH\endcsname &%
    \csname #1Wins#2#3StatecoverageFiveH\endcsname &%
    \num[round-pad=false]{\csname #1MeanObjectiveCoverage#2StatecoverageFiveH#3\endcsname}%
    \\}
    \newcommand{\createTableHeader}[0]{%
        & \multicolumn{2}{c}{Branch Coverage \%} & \multicolumn{2}{c}{Wins} & \multicolumn{1}{c}{MC \%}\\%
        \cmidrule{2-3} \cmidrule{4-5} \cmidrule{6-6}
        Project & \multicolumn{1}{c}{C} & \multicolumn{1}{c}{M} & C & M & M\\%
        }
    \newcommand{\createStatisticsRow}[0]{%
        Average &
        \num{\fpeval{\BcovcodeSum / \value{NumTableEntries}}} &
        \num{\fpeval{\BcovmodelSum / \value{NumTableEntries}}} \makebox[\widthof{$(\mathbf{0.00})$}][c]{$\boldsymbol{-}$} &
        \num{\fpeval{\codeWinsSum / \value{NumTableEntries}}} &
        \num{\fpeval{\modelWinsSum / \value{NumTableEntries}}} &
        \num{\fpeval{\mcovSum / \value{NumTableEntries}}} \\
    }
    \centering
    \caption{Average branch coverage, average model coverage (MC), effect size in parentheses, and number of wins for code-guided \Neatest (C) and model-guided \Neatest (M) using simple models. Bold indicates statistical significance.}
    \resizebox{\linewidth}{!}{
    \begin{tabular}[t]{l r r@{\extracolsep{10pt}} r r r@{\hspace{3pt}}|@{\hspace{3pt}}}
        \toprule
        \createTableHeader
        \midrule
        \createTableRow{CatchTheDots}{PlayThirtySecsimple}
        \createTableRow{CreateYourWorld}{Simple}
        \createTableRow{Dodgeball}{Simple}
        \createTableRow{Dragons}{Playfifteensecssimple}
        \createTableRow{FinalFight}{Win}
        \createTableRow[sprdb]{FlappyParrot}{ScoreOne}
        \createTableRow{Frogger}{Eatfly}
        \bottomrule
    \end{tabular}%
    \begin{tabular}[t]{l r r@{\extracolsep{10pt}} r r r}
        \toprule
        \createTableHeader
        \midrule
        \createTableRow{FruitCatcher}{SurviveThirtySecSimple}
        \createTableRow{HackAttack}{ScoreThirtySimple}
        \createTableRow[sprdb]{Pong}{SurviveFifteenSec}
        \createTableRow{Snake}{Scoreone}
        \createTableRow[sprdb]{SpaceOdyssey}{PlayThirtySec}
        \createTableRow{WhackAMole}{ScoreThirtysimple}
        \createStatisticsRow
        \bottomrule
    \end{tabular}
    }
    \setcounter{NumTableEntries}{0}
    \label{tab:basic_results}
    \vspace{-1em}
\end{table}

The achieved model coverage can be used to assess how well the trained neural network is able to achieve the task encoded in the model. For simple two-state models, the model coverage will always be either 0\% or 100\% in a single experiment repetition since there is only one state to reach. Thus, values between 0\% and 100\% indicate that the given task was achieved in some, but not all, experiment repetitions. In the \emph{CatchTheDots} game, the game model's final state was too hard to reach, as playing the game for 30 seconds proved impossible. However, branch coverage improves significantly if we guide the search towards networks that manage to play the game for a relatively long duration without losing. For the \emph{Dragons} game, the test generator achieved 100\% model coverage, but the branch coverage remains low, indicating that the task encoded in the model does not align with gameplay behaviour that maximises branch coverage.

\Cref{tab:basic_results} shows that code-guided \Neatest never manages to win the \emph{CreateYourWorld} game by reaching the next level, requiring the agents to touch the orange door depicted in \cref{fig:createyourownworld}. This
creates a deceiving fitness landscape, as the agents can only progress by temporarily moving away from the door and lowering their fitness score.
The model-guided approach always reaches the next level by no longer rewarding the agent for getting closer to the door, but improving fitness only if the agent reaches the next level. This turns the deceptive fitness landscape into a flat one, indicating that for this game, it is better for \Neatest to have no fitness guidance at all rather than a bad one.

RQ1 demonstrates that model-guided fitness
allows developers to focus the search on training agents that actually learn to play the game well rather than solely maximising coverage. Finally, the model-guided approach facilitates the design of fitness functions that help the search achieve the given task.

\subsection{RQ2 (Specialised Models)}
Since RQ1 showed that the model-guided approach assists the search, RQ2 investigates modelling techniques that further improve these reward signals. To this end, we select a subset of the models used in RQ1 and extend them with two techniques that generalise over game genres: \emph{Curriculum Learning} and \emph{sink states}. We focus on the games with worse performance in RQ1, for which the model-guided approach yields an average branch coverage of 76\%.
Using more specialised game models, however, we can increase the achieved branch coverage up to an average of 89\%.

The results of these experiments are shown in \Cref{tab:better_model_results}. While the simple models yield mixed results, the specialised models achieve notably better branch coverage than the code-guided approach by just playing the game better, even without specifically optimising for branch coverage.

\paragraph{Curriculum Learning:} The curriculum learning approach achieves higher coverage, since it tends to get stuck later in the learning process than the simple game models, while the more focused nature of the simple model does lead them to win the game slightly more often. This stepwise model progression of \emph{Curriculum Learning} is especially helpful for mastering games with lots of randomised behaviour, since it makes the search favour agents that can reliably progress in games rather than those that got lucky in specific program executions.

\begin{table}[t]
    \renewcommand{\BcovcodeSum}{0}
    \newcommand{\BcovModelSimpSum}{0}
    \newcommand{\BcovModelRefinedSum}{0}
    \newcommand{\CodeWinsSum}{0}
    \newcommand{\ModelWinSimpsSum}{0}
    \newcommand{\ModelWinsRefinedSum}{0}
    \renewcommand{\mcovSum}{0}
    \newcommand{\createTableRowNoStatistics}[5]{%
        #5%
        & \num{\csname #1MeanBranchCoverage#2StatecoverageFiveH#3\endcsname}%
        & \num{\csname #1MeanBranchCoverage#2StatecoverageFiveH#4\endcsname}\;\statisticsbox{\csname #1EffectSizeBranchCoverage#2NeatestBranchFiveHStatecoverageFiveH#4\endcsname}%
        & \csname #1Wins#2#3StatecoverageFiveH\endcsname%
        & \csname #1Wins#2#4StatecoverageFiveH\endcsname %
        & \num{\csname #1MeanObjectiveCoverage#2StatecoverageFiveH#4\endcsname}%
        \\
    }
    \newcommand{\createTableRow}[5][]{%
    \xdef\BcovcodeSum{\fpeval{\BcovcodeSum + \csname #1MeanBranchCoverage#2NeatestBranchFiveHNA\endcsname}}%
    \xdef\BcovModelSimpSum{\fpeval{\BcovModelSimpSum + \csname #1MeanBranchCoverage#2StatecoverageFiveH#3\endcsname}}%
    \xdef\BcovModelRefinedSum{\fpeval{\BcovModelRefinedSum + \csname #1MeanBranchCoverage#2StatecoverageFiveH#4\endcsname}}%
    \xdef\CodeWinsSum{\fpeval{\CodeWinsSum + \csname #1Wins#2NeatestBranchFiveH\endcsname}}%
    \xdef\ModelWinSimpsSum{\fpeval{\ModelWinSimpsSum + \csname #1Wins#2#3StatecoverageFiveH\endcsname}}%
    \xdef\ModelWinsRefinedSum{\fpeval{\ModelWinsRefinedSum + \csname #1Wins#2#4StatecoverageFiveH\endcsname}}%
    \xdef\mcovSum{\fpeval{\mcovSum + \csname #1MeanObjectiveCoverage#2StatecoverageFiveH#4\endcsname}}%
    \stepcounter{NumTableEntries}%
    \createTableRowNoStatistics{#1}{#2}{#3}{#4}{#5}
    }
    \newcommand{\createStatisticsRow}[0]{%
        Average &
        \num{\fpeval{\BcovModelSimpSum / \value{NumTableEntries}}} &
        \num{\fpeval{\BcovModelRefinedSum / \value{NumTableEntries}}} \makebox[\widthof{$(\mathbf{0.00})$}][c]{$\boldsymbol{-}$} &
        \num{\fpeval{\ModelWinSimpsSum / \value{NumTableEntries}}} &
        \num{\fpeval{\ModelWinsRefinedSum / \value{NumTableEntries}}} &
        \num{\fpeval{\mcovSum / \value{NumTableEntries}}}\\
    }
    \newcommand{\createTableHeader}[0]{
    & \multicolumn{2}{c}{Branch Coverage \%} & \multicolumn{2}{c}{Wins} & \multicolumn{1}{c}{MC \%}\\
        \cmidrule{2-3} \cmidrule{4-5} \cmidrule{6-6}
        Project & \multicolumn{1}{c}{RQ1} & \multicolumn{1}{c}{RQ2} & RQ1 & RQ2& \multicolumn{1}{c}{RQ2} \\
    }
    \centering
    \caption{Average branch coverage, average model coverage (MC), effect size in parenthesis and number of wins achieved for model-guided \Neatest with varying game models (RQ1/2). Values in bold indicate statistical significance. Only the best model for each game is used in the average.}
    \setlength{\tabcolsep}{0.8\tabcolsep}
    \resizebox{\linewidth}{!}{
    \begin{tabular}[t]{l r r@{\extracolsep{6pt}}r@{\hspace{-4pt}}r@{\hspace{-2pt}} r@{\hspace{3pt}}|@{\hspace{3pt}}}
        \toprule
        \createTableHeader
        \midrule
        \createTableRow{CatchTheDots}{PlayThirtySecsimple}{PlaySixtySec}{CatchTheDots}
        \createTableRowNoStatistics{}{CreateYourWorld}{Simple}{Es}{CYW\textunderscore path}
        \createTableRow{CreateYourWorld}{Simple}{EsCoverage}{CYW\textunderscore sidegoals}
        \createTableRow{Dodgeball}{Simple}{Dodgeball}{Dodgeball}
        \createTableRow{Dragons}{Playfifteensecssimple}{ScoreSixLongchain}{Dragons}
        \bottomrule
    \end{tabular}%
    \begin{tabular}[t]{l r r@{\extracolsep{6pt}}r@{\hspace{-4pt}}r@{\hspace{-2pt}} r}
        \toprule
        \createTableHeader
        \midrule
        \createTableRow[sprdb]{FlappyParrot}{ScoreOne}{ScoreSixLongchain}{FlappyParrot}
        \createTableRow{FruitCatcher}{SurviveThirtySecSimple}{ChainedSurvive}{FruitCatcher}
        \createTableRow[sprdb]{SpaceOdyssey}{PlayThirtySec}{CatchSevennolose}{SpaceOdyssey}
        \createTableRow{WhackAMole}{ScoreThirtysimple}{ScoreThirty}{WhackAMole}
        \createStatisticsRow
        \bottomrule
    \end{tabular}
    }
    \label{tab:better_model_results}
\end{table}

\begin{figure}[t]
    \begin{subfigure}{.32\linewidth}
        \includegraphics[width=\linewidth]{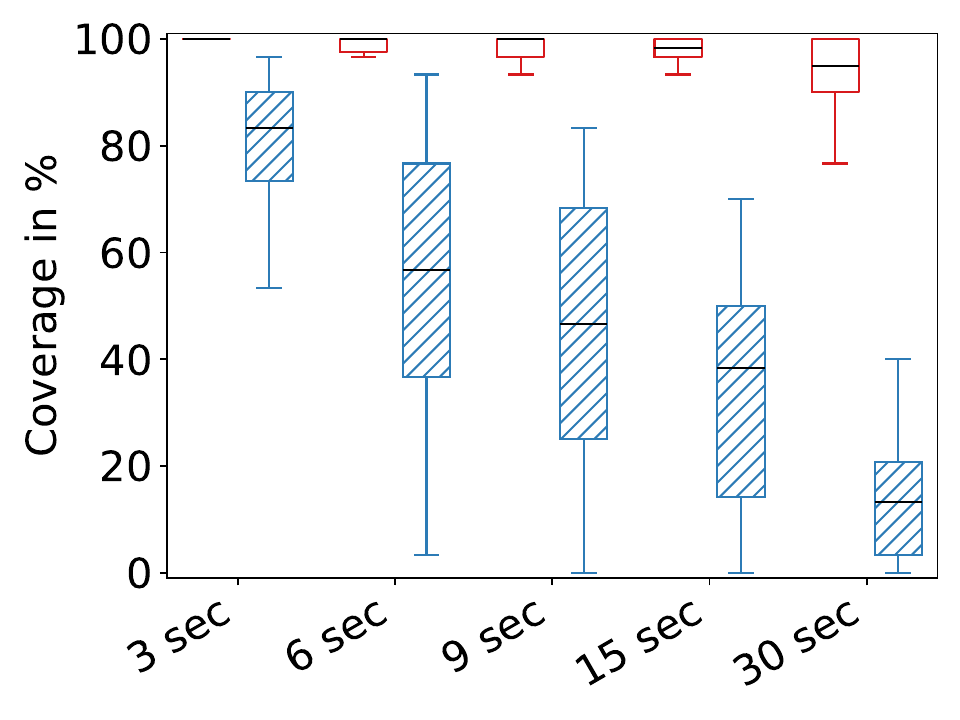}
        \caption{Pong: Play for 30s}
        \label{fig:pong_30sec}
    \end{subfigure}
    \begin{subfigure}{.32\linewidth}
        \includegraphics[width=\linewidth]{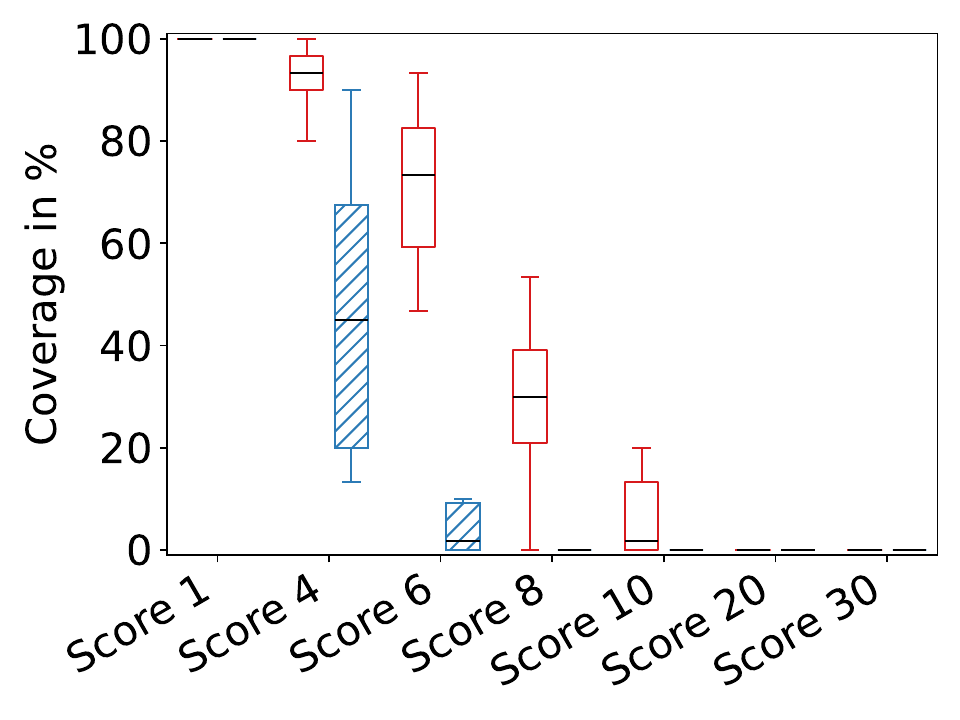}
        \caption{WhackAMole: Score 30}
        \label{fig:whackamole_score30}
    \end{subfigure}
    \begin{subfigure}{.32\linewidth}
        \includegraphics[width=\linewidth]{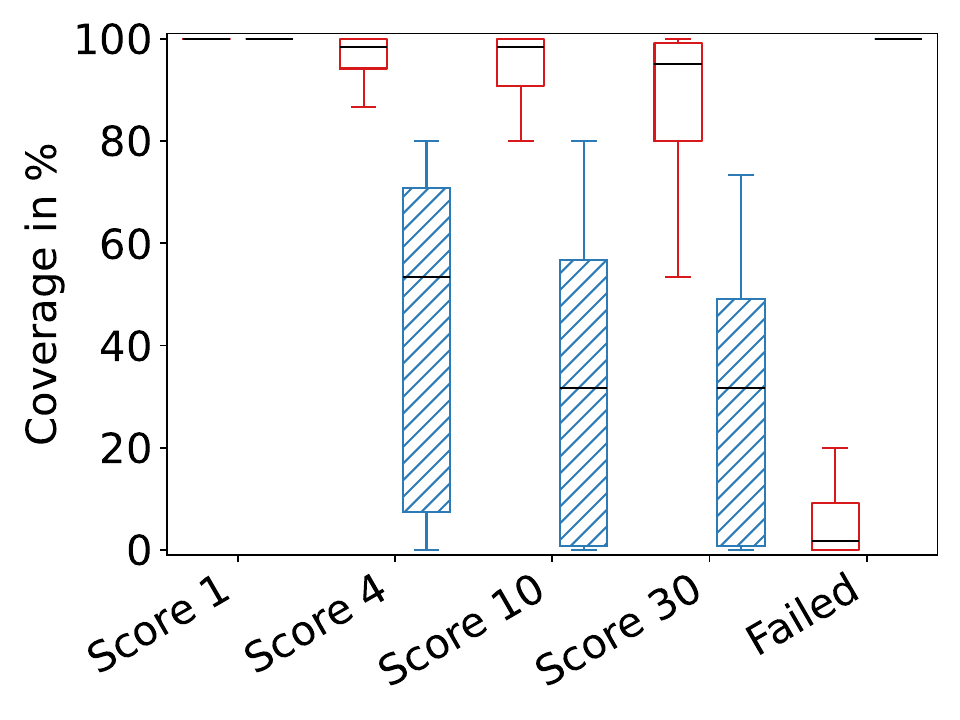}
        \caption{HackAttack: No damage}
        \label{fig:hackattack_play_well}
    \end{subfigure}
    \caption{Achieved state coverage using model-guided (red open) vs code-guided (blue striped) fitness functions across three different games.}
    \label{fig:statecoverage}
    \vspace{-1em}
\end{figure}

Comparing the results of the simple RQ1 model with those of the \emph{CYW\_path} model in \cref{tab:better_model_results} shows that adding intermediate waypoint states to the model improves the achieved branch coverage significantly from 74\% to 77\%. By adding waypoints, we can also encode side-goals
such as collecting the coin in \emph{CYW}, depicted in \cref{fig:createyourownworld}. These additional waypoints defined in the \emph{CYW\_sidegoals} models further improve the average coverage achieved from 77\% to 79\%.

\paragraph{Sink States:} The effects of adding sink states to the models are evident in \emph{CYW}. In this game, a playthrough ends if the player gets hit by a guard. However, since these guards only start appearing in later levels, every run that does not manage to reach the second level always uses the entire search budget for that run, even if it is not getting closer to the goal at all. By adding transitions to a sink state if the next goal is not reached within a customisable time limit, as depicted in \cref{fig:sink_chain_model}, we can stop futile agent evaluations early to save some time for more promising agents. This restriction not only improves search efficiency by reducing each agent's playtime but also allows us to define the agent behaviour, as the evolved agents must accomplish their task without reaching the sink state. For \emph{CYW}, the allocated search budget of 5 hours allows the code-guided \Neatest approach to evaluate, on average, 1800 agents and to evolve 12 generations. Using a model with sink states, we can significantly increase the number of evaluations and evolved generations to 14000 evaluations and 93 generations.

\paragraph{Targeted Coverage:} Both \emph{Pong} and \emph{WhackAMole} achieve the same branch coverage, regardless of whether the search is guided by a coverage- or model-derived fitness function, suggesting that the resulting test suites are equally suited to serve as test input generators for the game. However, \cref{fig:pong_30sec,fig:whackamole_score30} show that test suites generated by the model-based approach achieve higher state coverage with lower variance, indicating that they are better suited at testing the desired aspects of the game while also being more robust to random perturbations within the stochastic game environment.

Optimising test suites via a game model also allows testers to define specific gameplay strategies that ought to be tested, such as playing the game perfectly without losing a single life. For instance, \cref{fig:hackattack_play_well} shows the achieved state coverages on a \emph{HackAttack} model with a sink state that is entered whenever a single live is lost. As the results demonstrate, test suites generated via the model-guided search are better suited to exhibit the desired playstyle than coverage-based tests.

In the future, we want to explore using model-based tests to express concepts from traditional testing, such as equivalence partitioning, by defining several high-level strategies into different models and then letting the search algorithm find a test implementation that satisfies each.

\paragraph{Limitations:}
The game \emph{Dodgeball} starts by requiring the player to move with the arrow keys, but after some time it also tasks the player with jumping over obstacles by pressing the space bar. Such gameplay changes between states of the game model require deeper neural network models than those generated by \Neat after 5 hours of search. Alternatively, one could change the search to learn one network controller for each edge transition, rather than optimising agents that need to reach the targeted program state from the initial state.
For the \emph{Dragons} game, the lower performance can be explained by the large number of side conditions that are not required for winning, such as pressing all combinations of two arrow keys at once. For
\emph{SpaceOdyssey}, the problem is that the game relies heavily on sprite collisions with distorted sprite appearances, which our model abstraction does not properly handle due to limited information on the  \Scratch program state. In future work, we intend to address this by extending \Whisker's instrumentation of the \Scratch virtual machine.

Overall, RQ2 shows that the model-guided approach enables users to improve the search performance using sink states and \emph{Curriculum Learning} techniques. Furthermore, it allows testers to precisely define the behaviour that evolved agents should express, regardless of the game implementation, which is crucial for optimising automated playtesting agents that replicate varying testing scenarios.

%% file: sections/5-RelatedWork.tex
\section{Related Work}
Since its introduction, \Neatest has been extended with several techniques to improve its search efficiency, such as many-objective search~\cite{feldmeier_ManyObjectiveNeuroevolutionTesting_2025}, novelty search~\cite{feldmeier_CombiningNeuroevolutionSearch_2024}, and guided network weight optimisation via gradient descent using player traces~\cite{feldmeier_LearningViewingGenerating_2023}. 
Since all of these modifications to the \Neatest algorithm can be combined with any fitness function, they can also be applied to the model-guided optimisation framework presented in this paper. To allow for a better comparison between the code-guided and model-guided searches, we did not apply these extensions in this work and leave them for future work.

Previous work used models to derive test inputs for platform games~\cite{binder_TestingObjectorientedSystems_2000}, smoke test MMORPG quests~\cite{hu_LanguageguidedAccelerationMethod_2024}, solve levels in a stealth game~\cite{gutierrez-sanchez_ProgressBasedAlgorithmInterpretable_2024}, and test Android games~\cite{liang_AG3AutomatedGame_2023}. However, unlike our method these approaches are limited to specific game genres or test scenarios. Prior work uses \emph{Q-Learning}~\cite{sutton_ReinforcementLearningIntroduction_2018} to test games based on a model~\cite{prasetya_ModelBasedTestingComputer_2025}. However, in contrast to our method, the Q-Learning-based technique uses high-level input actions that require a navigation map, which significantly simplifies the search task but requires manual overhead and might not reflect how players would play the game. Overall, to the best of our knowledge, no testing framework exists that allows testers to create test scenarios for any game that are then realised by agents robust to program randomisation.

%% file: sections/6-Conclusions.tex
\section{Conclusions}
As generating tests for non-deterministic applications, such as games, is challenging, \Neatest generates test suites consisting of neural networks that dynamically adapt to changes in program behaviour. However, \Neatest suffers from scalability issues because it generates one test case for every branch or statement with no good way to focus the search. To overcome these issues, we combine \Neatest with a model-based testing framework, which allows game testers to define testing scenarios via abstract game models for which \Neatest then generates a network that replicates the desired gameplay behaviour. The model-guided approach evolves one agent per model state rather than one agent per branch. Since there are usually far fewer model states for a given testing scenario than there are branches in a program, the proposed approach scales much better to more complex programs. By using small, high-level models, we can keep the extra modelling effort low; by applying simple modelling techniques to specialise  models for gameplay challenges, we can increase the achieved coverage.

In future work, we aim to combine the model-guided approach with other \Neatest extensions, such as many-objective optimisation, to target several program states or  testing scenarios simultaneously. The neuro-evolutionary approach used could also be combined with more traditional model-based testing as a source of access sequences for hard-to-reach states or used as test oracles driven by reachability analysis, similar to previous work~\cite{gutierrez-sanchez_ProgressBasedAlgorithmInterpretable_2024}. Finally, we aim to apply the model-guided \Neatest framework in other application contexts with heavy program randomisation, such as autonomous driving or robot control.